\documentclass[twocolumn,superscriptaddress,
nofootinbib,
floatfix,
aps,
prb,preprintnumbers]{revtex4-2}
\pdfoutput=1
\usepackage{hyperref}
\usepackage{amsmath}
\usepackage{amsthm}
\usepackage{amsfonts}
\usepackage{amssymb}
\usepackage{mathtools}
\usepackage{textcomp}
\usepackage{graphicx}
\usepackage{bm}
\usepackage{color}
\usepackage{verbatim}
\usepackage{subcaption}
\usepackage{graphicx,color}
\usepackage{epsfig}
\usepackage{ulem}

\usepackage{cancel}

\definecolor{orange}{rgb}{1,0.5,0}
\definecolor{col1}{RGB}{153, 52, 121}
\definecolor{dgreen}{rgb}{0,0.55,0}
\definecolor{pink}{rgb}{1,0.08,0.58}

\usepackage{amsfonts,amsmath,amssymb}

\newcommand{\p}{\partial}

\newcommand{\la}{\langle}
\newcommand{\ra}{\rangle}

\newcommand{\rar}{\rightarrow}

\theoremstyle{definition}

\begin{document}


\title{Thermoelectric transport properties of gapless pinned charge density waves}
 
 \author{Tomas Andrade}
 \affiliation{
Departament de F{\'\i}sica Qu\`antica i Astrof\'{\i}sica, Institut de
Ci\`encies del Cosmos, Universitat de
Barcelona, \\ Mart\'{\i} i Franqu\`es 1, E-08028 Barcelona, Spain}
 
 \author{Alexander Krikun\footnote{https://orcid.org/0000-0001-8789-8703}}
 \affiliation{Nordita, 
KTH Royal Institute of Technology and Stockholm University \\
Hannes Alfvéns väg 12, SE-106 91 Stockholm, Sweden}

\preprint{NORDITA 2022-014}


\begin{abstract}
Quantum strongly correlated matter exhibits properties which are not easily explainable in the conventional framework of Fermi liquids. Universal effective field theory tools are applicable in these cases regardless of the microscopic details of the quantum system, since they are based on symmetries. It is necessary, however, to construct these effective tools in full generality, avoiding restrictions coming from particular microscopic descriptions which may inadequately constrain the coefficients that enter in the effective theory. 

In this work we demonstrate on explicit examples how the novel hydrodynamic coefficients which have been recently reinstated in the effective theory of pinned charge density waves (CDW) can affect the phenomenology of the thermo-electric transport in strongly correlated quantum matter. Our examples, based on two classes of holographic models with pinned CDW, have microscopics which are conceptually different from  Fermi liquids. Therefore, the above novel transport coefficients are nonzero, contrary to the conventional approach. We show how these coefficients allow to take into account the change of sign of the Seebeck coefficient and the low resistivity of the CDW phase of the cuprate high temperature superconductors, without referring to the effects of Fermi surface reconstruction.    
\end{abstract}

\maketitle

\section{Introduction}
The transport properties of the charge density wave (CDW) ordered states of strongly correlated quantum matter, in particular in cuprate superconductors, attract a great deal of attention. 
The CDW plays an important role in the phase diagram of the cuprates: it appears as an order parameter in a subregion across the pseudogap and superconductor phases, as well as a fluctuation at higher temperatures in the strange metal phase \cite{arpaia2021charge}.
It can be demonstrated \cite{collignon2021thermopower} that the onset of CDW leaves an imprint on the thermopower (or Seebeck coefficient), which starts decreasing at the CDW critical temperature $T_{\mathrm{CDW}}$ and reaches negative values at low $T$'s. This feature is observed in various compounds \cite{huker1998baberski,nakamura1992anisotropic,li2007two,chang2010nernst,doiron2013hall,badoux2016critical} and can be considered as a universal signature of CDW. 
From the point of view of the Fermi liquid theory, this behavior may be understood as a reconstruction of the Fermi surface \cite{laliberte2011fermi}, yet to be detected. However, the very applicability of the Fermi liquid to cuprates is debatable \cite{keimer2015quantum,reber2019unified,ayres2021incoherent} and the need of alternative approaches to the physics of thermopower in these systems persists \cite{gourgout2021out,georges2021skewed}.

Irrespective of the microscopic description of a particular system, one can construct an effective theory (EFT) of its low energy properties by symmetry considerations \cite{chaikin2000principles}. For exact global symmetries, this procedure leads to hydrodynamics. If a global symmetry is spontaneously broken, the corresponding Goldstone mode appears in the spectrum and must be added to the hydrodynamic degrees of freedom. 
If the spontaneously broken symmetry is only approximate (there is a small explicit breaking source), the hydrodynamic description of massive pseudo-Goldstone mode can be developed \cite{Delacretaz:2016ivq, Delacretaz:2017zxd, Delacretaz:2021qqu,Armas:2021vku}. The phases of matter with pinned CDW fall in the latter class: the spatial structure of CDW breaks translation symmetry spontaneously, while pinning provides an explicit symmetry breaking source.  

The effective theory of broken translations has long been appreciated as a tool to address the dynamics of the CDW phases in quantum systems \cite{RevModPhys.60.1129}. Recently, it has experienced a renaissance motivated in part by the appearance of a novel class of physical models described by the holographic duality \cite{Zaanen:2015oix,Hartnoll:2016apf}. These models, on one hand, can exhibit the CDW phases \cite{Amoretti:2017frz, Gouteraux:2018wfe,Donos:2018kkm,Amoretti:2019cef} (see \cite{Baggioli:2022pyb} for recent review) and therefore must be describable by a symmetry based EFT. On the other hand, they defy the principles of gapped Fermi liquid used in the earlier EFT constructions to constrain certain hydrodynamic coefficients. 
Therefore, the need to relax some of these constraints has been identified recently, leading to a new generation of EFT, with an enlarged set of nonzero hydrodynamic coefficients: the Galilean symmetry constraints have been relaxed in \cite{Delacretaz:2016ivq,Delacretaz:2017zxd,Delacretaz:2021qqu}, the effects of pinning where included in \cite{Delacretaz:2016ivq,Delacretaz:2017zxd,Delacretaz:2021qqu,Armas:2021vku} and the effects of background strain have been addressed in \cite{Armas:2019sbe,Armas:2020bmo}.

Here we study explicit examples of translational symmetry breaking in holographic models.
We evaluate all AC and DC conductivities and match the results with the EFT description of \cite{Armas:2020bmo,Delacretaz:2021qqu,Armas:2021vku}.
We show that the novel transport coefficients, namely, the incoherent conductivity and Goldstone mode diffusivity, are nonzero in the considered examples and lead to novel phenomenology: the change of sign of the Seebeck coefficient discussed above, and the absence of an exponential gap in the resistivity of the symmetry broken state. The latter is also a feature of CDW phase of cuprates \cite{takagi1992systematic,ando1995logarithmic,PhysRevMaterials.2.024804,laliberte2016origin}. 

In the following Section we review the hydrodynamic EFT of pinned CDWs. Then, we construct the holographic dual to a quantum system with CDW order and show that its AC conductivity, as expected, is very well described by the EFT, albeit with parameters which do not follow the conventional Fermi liquid logic. 
We obtain the DC conductivities and demonstrate some novel phenomenological features, which may be studied experimentally. 
The extra details are summarized in the Appendix (see, also, references \cite{deHaro:2000vlm,Withers:2013loa} therein).

\section{\label{sec:hydro}Effective theory of charge density waves}




%
In a hydrodynamic approach, the effective theory of pinned CDW can be built using exclusively the symmetry considerations 
\cite{Delacretaz:2016ivq,Delacretaz:2017zxd,Armas:2019sbe,Armas:2020bmo,Amoretti:2021fch,Armas:2021vku}, 
and, importantly, is valid irrespective of whether the quantum system admits a perturbative treatment or not.
Thus, EFT forms a convenient basis for the analysis of various experimental measurements, representing all data in terms of a limited number of hydrodynamic coefficients.

We will rely on the effective theory description of CDW developed in \cite{Armas:2019sbe,Armas:2020bmo,Amoretti:2021lll}. 
Let us restrict the model of \cite{Armas:2020bmo} to a space-time with 2 spatial dimensions ($x$-,$y$-).
with an unidirectional spontaneous spatial structure: CDW. Introducing an intrinsic coordinate $\phi$ along the CDW, one can characterize its embedding in space by a single ``crystal field'' $\phi(x,y)$. 
The ground state (a homogeneous CDW with no defects) corresponds to $\phi_0(x,y) = \alpha x$.
\footnote{$\alpha$ characterizes the ``wavelength'' of the CDW, however it is arbitrary given reparametrization freedom of $\phi$.} 
Note that the translation of the CDW as a whole -- the sliding mode -- is encoded in the shifts of $\phi$: $\phi \rar \phi + \delta \phi$ 
Therefore one readily identifies $\phi$ as a Goldstone field. 

We restrict our attention to small low-frequency, long-wavelength fluctuations of the CDW structure $\phi = \phi_0 + \alpha \delta \phi(t,x)$, the local chemical potential $\mu = \mu_0 + \delta \mu (t,x)$ and temperature $T=T_0 + \delta T (t,x)$, and local velocity field $u^\mu = \{1,0,0\} + \delta u^\mu(t,x)$. 
To compute the two-point functions relevant for transport, we consider external sources for the current and energy-momentum tensor: the electric field $\delta \p_t A_x(t,x)$ and the background metric perturbation $\delta g_{tx}(t,x)$. 
Moreover, we take into account the effect of the crystal lattice, which simultaneously breaks translations explicitly, introducing a finite momentum dissipation $\Gamma$, and pins the CDW, providing a mass $m_\phi^2$ to the Goldstone mode. 
As long as $\Gamma$ and $m_\phi^2$ are small, this ``weak pinning'' effect can be treated as a small correction to the hydrodynamic conservation laws. 
The pinning leads to a finite lifetime of the Goldstone, parametrized by the phase relaxation term $\Omega$ \cite{Delacretaz:2016ivq,Delacretaz:2017zxd,Andrade:2020hpu,Amoretti:2021fch,Amoretti:2021lll,Armas:2021vku,Delacretaz:2021qqu}.
The full set of hydrodynamic constitutive relations, resulting from the framework of \cite{Armas:2020bmo} is listed in Supplementatry material \cite{supp} Sec.\,A. Here we assume Lorentz symmetry \cite{Kovtun:2012rj} and the expression for the electric current and the Goldstone configuration equation (Josephson relation) read
\footnote{In \cite{Armas:2020bmo,Ammon:2020xyv} the extra thermodynamic quantity, ``lattice pressure'' plays a significant role in the model, but it is irrelevant here (see Supplementatry material \cite{supp} Sec.\,A)}
\begin{gather}
\label{equ:const_relations}
J^x \! = \rho \delta u^x  + \gamma (\p_t \delta \phi - \delta u^x) \! - \! \sigma_q \left( T_0 \p_x \frac{\mu}{T} + \p_t \delta A_x \right), \\
\label{equ:Josephson}
\begin{aligned}
\p_t \delta \phi - \delta u^x - &\frac{B + G}{\sigma_\phi} \p_x^2 \delta \phi \\ - &\frac{\gamma'}{\sigma_\phi} \left(T_0 \p_x \frac{\mu}{T} + \p_t \delta A_x \right) = -\Omega \delta \phi. 
\end{aligned}
\end{gather}
In addition, we quote the stress-energy and current conservation laws, modified by the explicit sources and symmetry breaking terms
\begin{align}
\nabla_\mu T^{\mu}_{t} &= 0, \qquad \nabla_\mu J^\mu = 0 \\
\nabla_\mu T^{\mu}_{x} &= - \rho \p_t  \delta A_x + \Gamma T_{t x} - G m_\phi^2 \delta \phi
\end{align}
Here $\nabla_\mu$ is a covariant derivative constructed with the perturbed metric, $\{P, \rho, s\}$ are thermodynamic pressure, charge density and entropy in the ground state, $\{B, G\}$ are bulk and shear elastic moduli of the spontaneous structure, $\{\zeta, \eta\}$ are bulk and shear viscosities, while $\{\sigma_q, \gamma, \gamma', \sigma_\phi\}$ are 4 more hydrodynamic coefficients which we discuss now in more detail.

%
%
The $\sigma_q$ and $\sigma_\phi^{-1}$ coefficients are usually set to zero in the standard treatment of CDW in gapped quantum systems\cite{RevModPhys.60.1129}. If we examine the Josephson relation in absence of perturbative sources and pinning, we see that at any finite momentum $k$ the Goldstone mode decays with the rate $\sigma_\phi^{-1}(B + G) k^2$, therefore $\sigma_\phi^{-1}$ controls the Goldstone diffusivity \cite{Amoretti:2018tzw,Amoretti:2019cef}. However in a gapped quantum system, the Goldstone mode cannot decay even at finite momentum, unless the momentum is large enough to cross the gap. Therefore, at zero temperature one sets $\sigma_\phi^{-1} = 0$ and all terms except $\delta u_x$ drop out from \eqref{equ:Josephson}. On the other hand, at finite temperature one expects $\sigma_\phi^{-1}$ to be exponentially suppressed by the scale of the gap. 

%
The other unusual coefficient is $\sigma_q$. This is allowed from the EFT perspective, but it is absent in systems where transport is mediated by quasiparticles and the current is constrained by Galilean symmetry $J^x = \rho \delta u^x$. This coefficient is related to ``incoherent conductivity'' which has been discussed extensively in connection to holographic models \cite{Davison:2015bea,Davison:2015taa, Amoretti:2017frz, Davison:2018ofp,Davison:2018nxm, Gouteraux:2018wfe,Donos:2018kkm,Amoretti:2019cef} and plays a crucial role here.   

We can plug in the constitutive relations into the conservation laws and solve the system of differential equations with respect to hydrodynamic variables $\{\delta u^x, \delta \mu, \delta T, \delta \phi\}$ in presence of external sources $\{\delta A_x, \delta g_{tx}\}$. Inserting the solutions back into the constitutive relations, we obtain the expectation values for various operators in terms of perturbative sources \cite{Kovtun:2012rj}. This allows us to evaluate the two point functions $\la J^x J^x \ra$, $\la J^x T^{tx} \ra$, $\la T^{tx} J^x \ra$, and $\la T^{tx} T^{tx} \ra$. As one can show from locality of the hydrodynamic equations\cite{Delacretaz:2021qqu}, positivity of entropy production \cite{Armas:2021vku}, or the Onsager relation between $\la J^x T^{tx} \ra$ and $\la T^{tx} J^x \ra$ (see Appendix Sec.\,A) the coefficients are related as\cite{Amoretti:2021fch,Amoretti:2021lll,Amoretti:2018tzw,Ammon:2019wci,Donos:2019hpp,Amoretti:2019cef,Andrade:2018gqk, Baggioli:2020nay, Baggioli:2020haa}.
\begin{equation}
\label{equ:Onsager}
\gamma' = - \gamma, \qquad \Omega = \sigma_\phi^{-1} m_\phi^2 G.
\end{equation}

Recalling the definition of the heat current in presence of a chemical potential \cite{Herzog:2009xv}: $Q^x = J^x + \mu T^{t x}$, we arrive at the full matrix of AC thermoelectric conductivities \cite{Hartnoll:2009sz} ($\bar \kappa$ is thermal conductivity at zero bias)
\begin{equation}
\label{eq:sigma_matrix}
\begin{pmatrix}
J^x \\ Q^x
\end{pmatrix} = 
\begin{pmatrix}
\sigma & T \alpha \\ T \bar \alpha & T \bar \kappa
\end{pmatrix}
 \begin{pmatrix}
E_x \\ - \frac{\p_x T}{T}
\end{pmatrix}
\end{equation}.
At zero wavelength limit
\begin{align}
\notag
\sigma(\omega)\! &= \! \sigma_0 \! + \! \frac{\tilde{\rho}^2 (\Omega - i \omega) - \tilde{\gamma}^2 \omega_0^2 (\Gamma \! - \! i \omega) - 2 \tilde{\rho} \tilde{\gamma} \omega_0^2}{ \mu_0^2 \chi_{\pi \pi} ((\Gamma -  i \omega)(\Omega - i \omega) + \omega_0^2) } \\
\notag
\frac{T}{\mu_0} \alpha(\omega)\! &= \! - \! \sigma_0 \! + \! \frac{\tilde{\rho} \tilde{s} (\Omega \! - \! i \omega) + \tilde{\gamma}^2 \omega_0^2 (\Gamma \! - \! i \omega) - (\tilde{s} - \tilde{\rho}) \tilde{\gamma} \omega_0^2}{ \mu_0^2 \chi_{\pi \pi} ((\Gamma - i \omega)(\Omega - i \omega) + \omega_0^2) } \\
\label{equ:AC_formulae}
\frac{T}{\mu_0^2} \bar{\kappa}(\omega)\! &= \! \sigma_0 \! + \! \frac{\tilde{s}^2 (\Omega - i \omega) - \tilde{\gamma}^2 \omega_0^2 (\Gamma \! - \! i \omega) + 2 \tilde{s} \tilde{\gamma} \omega_0^2}{\mu_0^2 \chi_{\pi \pi} ((\Gamma - i \omega)(\Omega - i \omega) + \omega_0^2) }, 
\end{align}
where
\begin{gather}
\sigma_0 = \sigma_q + \frac{\gamma^2}{\sigma_\phi}, \ \ \tilde{\rho} = \mu_0 \rho, \ \ \tilde{s} = T_0 s, \ \ \chi_{\pi \pi} = \tilde{\rho} + \tilde{s} \\
\notag
\tilde{\gamma} = \mu_0 \chi_{\pi \pi} \gamma/\sigma_\phi, \ \ \omega_0^2 = G m_\phi^2/\chi_ {\pi \pi}.
\end{gather}
In the ordered state far from $T_c$ ($\omega_0 \gtrsim \Omega,\Gamma$) these expressions correspond to the peak in the real part of the spectra, located at finite ``pinning frequency'' $\omega_0$ with width $(\Omega + \Gamma)$. This is a manifestation of the gapped coherent sliding mode.
The DC conductivities display a mixture of coherent and incoherent contributions, which can be recast as
\begin{gather}
 \notag
\sigma = \sigma_q + \sigma_\phi^{-1} \frac{(\rho - \gamma)^2}{1 + \frac{\Gamma  \chi_{\pi \pi}}{\sigma_\phi}}, \quad  \frac{\bar{\kappa} T}{\mu^2} = \sigma_q + \sigma_\phi^{-1}  \frac{(\frac{s T}{\mu} + \gamma)^2}{1 + \frac{\Gamma  \chi_{\pi \pi}}{\sigma_\phi}},
 \\
\label{equ:DC_hydro}
\frac{\alpha T}{\mu} = - \sigma_q + \sigma_\phi^{-1} \frac{(\rho - \gamma)(\frac{s T}{\mu} + \gamma)}{1 + \frac{\Gamma  \chi_{\pi \pi}}{\sigma_\phi}}.
\end{gather}
These expressions represent the main outcome of the EFT\cite{Armas:2019sbe,Amoretti:2021fch,Delacretaz:2021qqu}, which we are going to study. Note that both terms are usually negligible in the conventional gapped CDW: $\sigma_q$ is zero because of Galilean invariance, while $\sigma_\phi^{-1}$ is exponentially suppressed due to a gap in the spectrum.

In presence of finite $\sigma_q$ and $\sigma_\phi^{-1}$, however, the DC conductivities depart from the conventional picture. Firstly, the electric conductivity is finite even in the broken phase, and secondly, the sign of the thermopower $\alpha$ is a result of the interplay between $\sigma_q$ and $\sigma_\phi^{-1}$ terms. In what follows we explore how this novel mechanisms come about in non-Fermi liquid, holographic models.

\section{\label{sec:model}Holographic model of gapless CDW}

The essence of holographic modelling is the correspondence between strongly correlated quantum systems and classical black holes in auxiliary spacetimes, which are constructed according to a rigorous set of rules --  the ``holographic dictionary'' \cite{Zaanen:2015oix,Hartnoll:2016apf,Herzog:2009xv,Hartnoll:2009sz}. In this paradigm, the quantum system at finite temperature $T$ and chemical potential $\mu$ in a 2+1 dimensional space-time corresponds to a black hole in a 3+1 dimensional curved space-time, whose horizon radius and charge are set by $T$ and $\mu$. The crystal lattice can be introduced via periodic modulation of the chemical potential \cite{Flauger:2010tv,Liu:2012tr,Horowitz:2012ky,Horowitz:2012gs,Donos:2014yya,Rangamani:2015hka,Langley:2015exa}: $\mu(x) = \mu_0 (1 + A \cos(k x))$ \footnote{This unidirectional crystal model allows us to simplify the treatment preserving all the necessary physics.}. We will consider a model with small $A = 0.04$, describing the weakly pinned CDW. The spontaneous structure formation is realized as an instability of the black hole against formation of the spatially modulated ``hair'' \cite{Donos:2013wia,Withers:2013kva,Donos:2013cka}.
%
The interplay between the explicit and spontaneous translation symmetry breaking has been studied extensively in this setup \cite{Krikun:2017cyw,Andrade:2017ghg,Andrade:2017leb,Andrade:2020hpu} (note also other approaches \cite{Andrade:2013gsa,Donos:2013eha,Baggioli:2014roa,Alberte:2015isw,Amoretti:2018tzw,Ammon:2019wci,Donos:2019hpp,Amoretti:2019cef,Baggioli:2022pyb,Baggioli:2021xuv,Nakamura:2009tf,Ooguri:2010kt,Donos:2012wi,Donos:2012js,Donos:2014oha,Andrade:2018gqk}, see Suplementary Material\,Sec.\,D.)
The action of the model reads
\begin{multline}\label{S_0}
  S =\!\int d^4\!x \sqrt{- g} \left(\!R\!-\!2 \Lambda- \frac{1}{2} (\partial \psi)^2 - \frac{\tau(\psi)}{4} F^2 -  W(\psi) \right) \\
   - \frac{1}{2} \int {\theta}(\psi) F \wedge F.
\end{multline}
where $F=dA$ is the field strength of the $U(1)$ gauge field dual to a $U(1)$ global charge, $\psi$ is an axion field in the bulk coupled to the $\theta$-term, which drives a CDW instability. $R$ and $\Lambda$ are the Ricci curvature and negative cosmological constant, which govern the structure of asymptotically Anti-de Sitter (AdS) space in the bulk. The qualitative features we reveal depend only mildly on the precise form of the potentials
\begin{gather}
\notag
\tau(\psi)= 1 + \dots, \qquad W(\psi) = - \psi^2 + \dots, \\
\label{eq:CScoupling}
 \theta(\psi) = \frac{c_1}{2 \sqrt{6}} \psi + \dots, 
\end{gather}
for more details see Supplementary Material \cite{supp} Sec.\,B.

The holographic dictionary identifies the asymptotics of the gauge field profile near the AdS boundary (located at radial coordinate $z \rar 0$) with the chemical potential and $U(1)$ charge density: $A_t(x,z)\Big|_{z\rar 0} = \mu(x) z + \rho(x) z^2$. Given the classical solution to the Einstein equations following from \eqref{S_0} with appropriate boundary conditions set by $\mu(x), T$ one can compute the charge density profile $\rho(x)$ and observe formation of spontaneous CDW below a certain critical temperature $T=T_0(c_1)$. 
As the temperature is lowered, the order parameter -- the amplitude of the charge density modulation
-- grows and the effective theory of pinned CDW is applicable. One can achieve a similar behavior tuning the coupling constant $c_1$ at fixed temperature.

To evaluate the AC conductivities, we introduce perturbative sources for the electric current and stress-energy tensor,  encoded in the near-boundary asymptotes of the gauge field $A_x(z)$ and metric $g_{tx}(z)$. After solving the equations of motion,
we read off the subleading components and take the variations with respect to the sources.
We thus obtain the 2-point functions $\la J^x J^x \ra$, $\la J^x T^{tx} \ra$, $\la T^{tx} J^x \ra$, and $\la T^{tx} T^{tx} \ra$ 
from which we read all AC thermoelectric conductivities \eqref{eq:sigma_matrix}. 
The numerical calculations are demanding, especially at low temperatures. To circumvent this, we fix the temperature and tune the coupling  $c_1$ to control the order parameter, see Appendix Sec.\,B for details. 

\begin{figure}
\includegraphics[width=1 \linewidth]{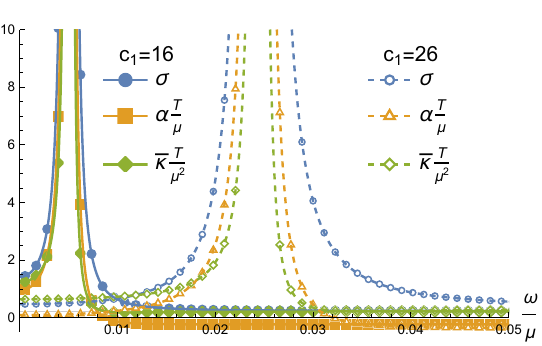}
\caption{\label{fig:AC_cond} The pinned peak in AC conductitivities. The filled dots and solid lines show the AC data for the conductivities in the holographic model with $c_1 = 16$ and the fits with \eqref{equ:AC_formulae}. The empty dots and dashed lines -- the same with $c_1=26$. Data taken for the $2/1$ commensurate states with $T/\mu=0.1, k/\mu\approx 2.1, A=0.04$.}
\end{figure}

In the weak pinning regime ($A= 0.04$), the results have precisely the shape predicted by the effective theory, see Fig.\ref{fig:AC_cond}: the peak located at finite frequency $\omega_0$. 
As we discuss in Appendix Sec.\,C, we perform a set of cross-checks of the model expressions \eqref{fig:AC_cond}. 
Firstly, we obtain the same values of hydrodynamic coefficients when independently fitting $\sigma, \alpha$ and $\bar \kappa$. 
Moreover, we extract thermodynamic data from the AC linear response fits, which agree with the data obtained as the operator expectation values in the ground state of the model. 
This check also shows that the extra thermodynamic quantities, discussed in \cite{Armas:2019sbe,Armas:2019sbe,Armas:2021vku,Ammon:2020xyv}, like the lattice pressure, can be safely neglected.   
Finally, we get excellent agreement between the DC conductivities, obtained by the expressions \eqref{equ:DC_hydro} using the values of the hydrodynamic coefficients from the AC fits, and the DC transport properties evaluated using the near horizon data in the holographic ground state, as we discuss in a moment.
These checks support the statement that the effective theory of pinned CDW from Sec.\,\ref{sec:hydro} describes well the holographic results.
\begin{figure*}[t]
\includegraphics[width=0.32 \linewidth]{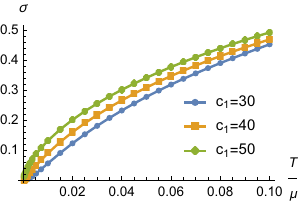} \
\includegraphics[width=0.32 \linewidth]{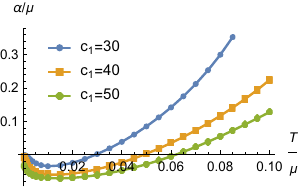} \ 
\includegraphics[width=0.3 \linewidth]{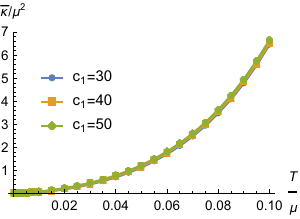} \ 
\caption{\label{fig:DCcond} The evolution DC thermo-electric conductivities at small temperatures. The holographic model with different values of the coupling $c_1$ is considered. Note that thermopower changes sign at some coupling-dependent point. Data taken for the $2/1$ commensurate states with $k/\mu = 2, A=0.04$.}
\end{figure*}

\section{\label{sec:DC}DC transport}

The DC transport of the holographic model can be studied with greater precision than the AC conductivities since it can be extracted
from near horizon data of the background geometry \cite{Donos:2018kkm,Donos:2017mhp,Donos:2015bxe,Banks:2015wha,Donos:2014cya,Donos:2014yya,Iqbal:2008by}. 

\begin{figure}[t]
\includegraphics[width=0.8 \linewidth]{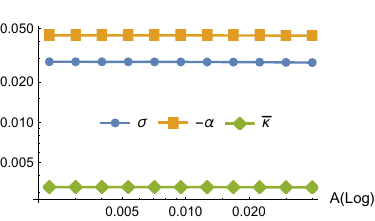}
\caption{\label{fig:DC_vs_A} Dependence of the DC thermo-electric conductivities on the pinning scale $A$. The data taken for the $2/1$ commensurate states with $k/\mu = 2, A=0.04$ }
\end{figure}

The results for temperature series of solutions with different $c_1$-coupling are shown on Fig.\,\ref{fig:DCcond}. One can recognize the unconventional behavior of the electric conductivity, which decreases at small temperature as a certain power law, instead of the activated exponential behavior ($\sim \exp(- \Delta_{CDW}/T)$) expected for the gapped CDW. 
Moreover the thermopower changes sign at a certain temperature, which depends on the coupling $c_1$. While being unusual in the conventional treatment, this behavior can be well incorporated in the generalized EFT framework, Sec.\ref{sec:hydro}. 
Note also that, as we show on Fig.\,\ref{fig:DC_vs_A}, the DC conductivities are insensitive to the scale of the explicit symmetry breaking (crystal lattice/impurities), which supports our treatment of $\Gamma$ as a small parameter in \eqref{equ:DC_hydro}.  

\begin{figure*}[t]
\includegraphics[width=0.32 \linewidth]{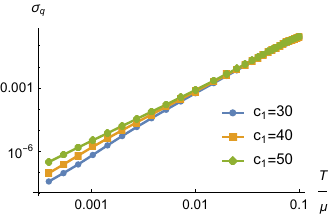} \
\includegraphics[width=0.32 \linewidth]{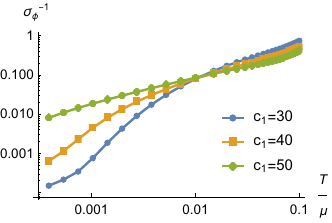} \
\includegraphics[width=0.32 \linewidth]{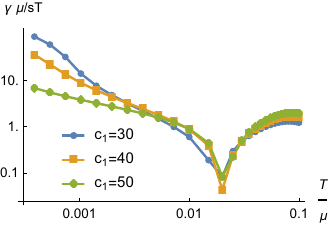}
\caption{\label{fig:EFT_vs_T} The behavior of effective theory parameters at low temperatures. Log-Log scales. Incoherent conductivity $\sigma_q$ and Goldstone diffusivity behave as certain power laws, but the former does not depend on the coupling $c_1$. $\gamma$ changes sign at some $T/\mu \approx 0.02$ for all $c_1$. Note that this is not related to the change of sign in $\alpha$. $\gamma \mu/ s T$ doesn't quite saturate at 1, as suggested in \cite{Amoretti:2019cef,Amoretti:2018tzw}. Same data as on Fig.\,\ref{fig:DC_vs_A}.}
\end{figure*}

In the previous section we established the validity of EFT as we vary the order parameter dialling $c_1$. Assuming that EFT continues to be applicable in the regimes where the order grows due to the decrease of temperature, we can use \eqref{equ:DC_hydro} and extract all the hydrodynamic coefficients having the DC conductivities at hand (in the weak pinning regime). This leads to the results shown on Fig.\,\ref{fig:EFT_vs_T}. We see that the EFT parameters $\sigma_q$ and $\sigma_\phi^{-1}$
are indeed non-zero and behave as power laws rather then as gapped exponentials. 

\begin{figure*}[t]
\includegraphics[width=0.32 \linewidth]{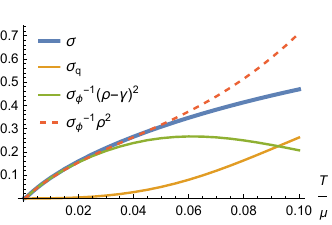} \
\includegraphics[width=0.32 \linewidth]{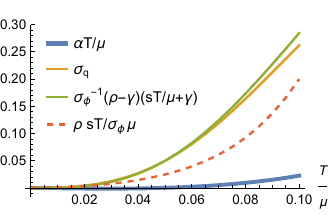} \
\includegraphics[width=0.32 \linewidth]{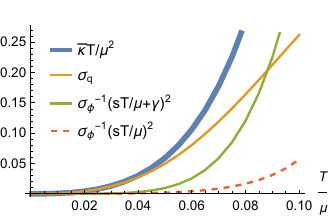}
\caption{\label{fig:EFT_contrib} The contributions to DC conductivities from various terms in \eqref{equ:DC_hydro}. At low $T$ electric conductivity is dominated by $\sigma_\phi^{-1}$ term, heat conductivity -- by $\sigma_q$-term, while thermopower is a result of a fine balance between them. Red lines assume $\gamma=0$ and help to appreciate the role of $\gamma$. Data as on Fig.\,\ref{fig:DC_vs_A} for $c_1=40$.}
\end{figure*}

The relative contributions of the various terms in \eqref{equ:DC_hydro} to the DC transport are shown on Fig.\,\ref{fig:EFT_contrib}. Interestingly, we see that the electric conductivity is dominated by the $\sigma_\phi^{-1} \rho^2$ term, while the heat conductivity ($\bar \kappa$ and, as we checked,  $\kappa = \bar \kappa - \alpha^2/\sigma$) is controlled by $\sigma_q$. This means that the behavior of the electric and heat conductivities is not related to each other at the level of EFT and the Wiedemann–Franz law ($\kappa/\sigma \sim T$) may not arise in materials with these ``gapless pinned CDW''.
Looking at the thermopower, we see that the two terms in \eqref{equ:DC_hydro} are of the same order and approximately cancel each other. This cancellation is the reason of the sign change in $\alpha$ as seen on Fig.\,\ref{fig:DCcond}. 
Therefore, the exact temperature where the sign changes depends on subleading contributions, and does not point to some qualitative change, as it would be for Fermi surface reconstruction. 

The phenomenological features which we observe here are not specific to a particular holographic model. As we show in Appendix Sec.\,D, we obtain similar results in a different setup based on a helical lattice. The universal feature, which appears in both cases is the absence of exponential suppression of either $\sigma_q$ or $\sigma_\phi^{-1}$, leading to a nontrivial interplay between the terms in \eqref{equ:DC_hydro}. 
The remarkable cancellation of terms in $\alpha$ as well as the exchange of dominance of the $\sigma_q$ and $\sigma_\phi^{-1}$ terms in $\sigma$ and $\bar \kappa$ are also persistent features in our data. This might point out some universal relation between $\gamma \sigma_\phi^{-1}$ and $\sigma_q$ as suggested in \cite{Amoretti:2019cef,Amoretti:2018tzw}, see Appendix Sec.\,E for the further discussion on this. 

\section{Conclusion}
In this Letter we demonstrate how the novel parameters in the effective theory of pinned charge density waves \cite{Delacretaz:2016ivq,Delacretaz:2017zxd,Armas:2019sbe,Armas:2020bmo,Amoretti:2021lll,Armas:2021vku,Delacretaz:2021qqu}, which are absent in gapped quantum systems, can affect the thermo-electric transport when they are nonzero. 
The absence of the clean gap in the spectrum can be a result of the quantum continuum being present in the system. The effects of such a continuum spectrum have been studied in relation to the physics of plasmons \cite{Krikun:2018agd}, the dynamics of the pattern formation \cite{Andrade:2020hpu} and fermion spectral functions \cite{Balm:2019dxk,Gnezdilov:2018qdu,smit2021momentum}.

In the explicit examples constructed by means of holographic duality, we show that the ``gapless'' pinned charge density waves demonstrate non-vanishing conductivity in the pinned CDW phase, change of sign in the thermopower, unrelated to the reconstruction of the Fermi surface, the conceptual absence of Wiedemann–Franz law 
and the low sensitivity of the transport properties to the concentration of impurities. All these phenomenological features are observed in the underdoped phases of cuprate high temperature superconductors, where charge density waves are present.

We hope that these examples will encourage the use of the improved EFT in the analysis of the transport experiments in CDW cuprates, which in turn could clarify their physical nature and, perhaps, would unveil the properties of the underlying quantum criticality.

We also point out that the phenomenology of the DC transport, described above, being a consequence of the pinned CDW behavior, predicts the specific shapes of the spectra of AC conductivities \eqref{equ:AC_formulae}, which makes the experiments on the optical spectroscopy particularly important.       

It is worth mentioning that when deriving the expressions \eqref{equ:DC_hydro} and \eqref{equ:AC_formulae} and focusing on the qualitative effects, we omitted other novel hydrodynamic coefficients, pointed out in \cite{Armas:2021vku}, which play a subleading role in our discussion. One should keep this in mind when performing precision tests of \eqref{equ:DC_hydro} and \eqref{equ:AC_formulae}.

Finally, we point out that the considered EFT framework resides on the Lorentz symmetry and therefore the ``incoherent contributions'' $\sim \sigma_q$ in all the conductivities \eqref{equ:DC_hydro} are related. In absence of Lorentz symmetry one allows instead 3 independent coefficients: $\sigma_q, \alpha_q$ and $\kappa_q$.
A nonzero negative $\alpha_q$ of order $\sigma_q$ could lead to the similar cancellation phenomenon in thermopower.

\begin{acknowledgments}
We thank Koenraad Schalm, Jan Zaanen and Floris Balm for long lasting collaboration in this subject. We are grateful to Jay Armas, Akash Jain, Andrea Amoretti, Daniel Brattan and Blaise Gouteraux for useful communication and valuable advice. A.K. thanks Dragana Popovic, Floriana Lombardi, Riccardo Arpaia, Ulf Gran and David Marsh for insightful comments.

A.K. acknowledges the hospitality of the physics department of Chalmers Technical University, where the preliminary results of this work have been discussed. 

The work of A.K is supported by VR Starting Grant 2018-04542 of Swedish Research Council.
%
The work of T.A. is supported in part by the ERC Advanced Grant GravBHs-692951 and by Grant CEX2019-000918-M funded by Ministerio de Ciencia e Innovaci\'on (MCIN)/Agencia Estatal de Investigaci\'on (AEI)/10.13039/501100011033.

The numerical computations were enabled by resources provided by the Swedish National Infrastructure for Computing (SNIC), partially funded by the Swedish Research Council through grant agreement no. 2018-05973, at SNIC Science Cloud and PDC Center for High Performance Computing, KTH Royal Institute of Technology.

Nordita is supported in part by Nordforsk.

\end{acknowledgments}

\appendix

\begin{widetext}

\section{\label{app:hydro}Effective theory of CDW}

In order to derive equations (7) in the main text we rely on the framework used in \cite{Armas:2019sbe} and supplement it with the explicit symmetry breaking terms introduced in \cite{Delacretaz:2017zxd}. We restrict the dynamics to a single $x-$ direction and consider a unidirectional CDW. The full set of hydrodynamic constitutive relations we get reads   

\begin{align}
\label{equ:full_const_relations}
J^t &= \rho + \left(\frac{\p \rho}{\p \mu} \delta \mu + \frac{\p \rho}{\p T} \delta T \right) + \frac{\p P_l}{\p \mu} \p_x  \delta \phi , \\
J^x &= \rho \delta u^x  + \gamma (\p_t \delta \phi - \delta u^x) - \sigma_q \left( T_0 \p_x \frac{\mu}{T} + \p_t \delta A_x \right), \\
T^{tt} &= \epsilon + \left(\mu_0 \frac{\p \rho}{\p \mu} + T_0 \frac{\p \rho}{\p T} \right) \delta \mu  + \left(\mu_0 \frac{\p s}{\p \mu} + T_0 \frac{\p s}{\p T} \right) \delta T + \left[-P_l + \mu_0 \frac{\p P_l}{\p \mu} + T_0 \frac{\p P_l}{\p T}  \right] \p_x \delta \phi,   \\
T^{tx} &= P \delta g_{tx} + (\mu_0 \rho + T_0 s) \delta u^x + P_l (\delta g_{tx} + \p_t \delta \phi), \\
T^{xx} &= P  + \Big( s \delta T + \rho \delta \mu + (\zeta + \eta) \p_x \delta u^x - (B+G) \p_x \delta \phi - 2 \eta \p_x \delta g_{tx} \Big) + \left[ P_l + \frac{\p P_l}{\p T} \delta T + \frac{\p P_l}{\p \mu} \delta \mu + P_l \p_x \phi \right],  \\
\p_t \delta \phi &- \delta u^x - \frac{B + G}{\sigma_\phi} \p_x^2 \delta \phi - \frac{\gamma'}{\sigma_\phi} \left(T_0 \p_x \frac{\mu}{T} + \p_t \delta A_x \right) - \Big[P_l (\p_x^2 \delta \phi + \p_t^2 \delta \phi + \p_t \delta g_{tx}) + \p_x P_l \Big]= -\Omega \delta \phi. 
\end{align}
Following the procedure outlined in the main text and taking the variations with respect to the sources $\delta g_{tx}$, $\delta A_x$, we arrive, among others, to the following expressions for the correlation functions
\begin{align}
\la J_x T_{tx} \ra &= - \left(\rho (\omega + i \Omega) + i \gamma' \sigma_\phi^{-1} \chi_{\pi \pi} \omega_0^2 \right)/ (\omega^2 + i \omega (\Gamma + \Omega) - \Gamma \Omega - \omega_0^2), \\
\la T_{tx} J_{x} \ra &= - \left(\rho (\omega + i \Omega) - i \gamma \Omega \right)/ (\omega^2 + i \omega (\Gamma + \Omega) - \Gamma \Omega - \omega_0^2), \\
\la T_{tx} T_{tx} \ra &= - \left(\chi_{\pi \pi} (\omega + i \Omega)\right)/ (\omega^2 + i \omega (\Gamma + \Omega) - \Gamma \Omega - \omega_0^2).
\end{align}

Where $P_l(T,\mu)$ is the ``lattice pressure'' introduced in \cite{Armas:2019sbe,Armas:2021vku,Ammon:2020xyv}. It must be zero for thermodynamically stable solutions, but can emerge in some holographic models \cite{Baggioli:2022pyb}. Its derivatives can be nonzero even for stable groundstates and appear in the AC 2-point functions. The complete expressions for AC thermoelectric transport coefficients can be found in \cite{Amoretti:2021lll}.

However the terms involving $P_l$ do drop out from the expressions of DC thermoelectric conductivities which we mainly focus on. We also check it from our fits of the AC data that the thermodynamic values extracted from the fit assuming $P_l \equiv 0$ match well with the data obtained directly from the 1-point functions, which demonstrates that $P_l$ plays subleading role at most. Therefore we drop these  terms from the equations throughout the text and obtain (7) and (8) of the main text from \eqref{equ:full_const_relations}, given the definition of the heat current.

\section{\label{app:periodic_numerics}Periodic holographic model}

In the main text we mostly rely on the holographic model with periodic ionic lattice \cite{Horowitz:2012gs} and include a spontaneous translation symmetry breaking mechanism first introduced in \cite{Donos:2011bh}. As outlined in the main text the action is  
\begin{equation}\label{S_0_a}
  S = \int d^4 x \sqrt{- g} \left( R - 2 \Lambda- \frac{1}{2} (\partial \psi)^2 - \frac{\tau(\psi)}{4} F^2 -  W(\psi) \right)
   - \frac{1}{2} \int {\vartheta}(\psi) F \wedge F.
\end{equation}
The potentials are
\begin{gather}
\label{equ:potentials}
  V(\psi) \equiv 2 \Lambda +W(\psi) = - 6 \cosh (\psi /\sqrt{3}) , \quad \\
  \notag
   \tau(\psi) = {\rm sech} (\sqrt{3} \psi), \quad \vartheta(\psi) = \frac{c_1}{6 \sqrt{2}} \tanh(\sqrt{3} \psi),
\end{gather}
Note that in these conventions the cosmological constant is $\Lambda = - 3$ and the mass of the scalar is $m^2 = -2$. This setup is identical to the one we used earlier in \cite{Krikun:2017cyw,Andrade:2017ghg,Andrade:2020hpu,Andrade:2017leb} and we refer the reader to these papers for the details of our calculation scheme. The solution to nonlinear equations of motion following from \eqref{S_0_a} can be found in the ansatz
\begin{align}\label{ds2 anstaz}
  ds^2 &= \! \frac{1}{z^2}\left( \! - Q_{tt} f(z) dt^2 + Q_{zz} \frac{dz^2}{f(z)} + Q_{xx} (dx + Q_{zx} dz)^2 + Q_{yy} ( dy + Q_{ty} dt )^2  \right), \\
  {\cal A} &= A_t dt + A_y dy, 
\end{align}
where $f = (1-z)\left( 1 + z + z^2 - \mu^2 z^3 /4 \right)$ and the temperature is related to the chemical potential as $T/\mu = (12 - \mu^2)/(16 \pi \mu)$. All the ansatz functions are periodic in $x$ and the period is set by the explicit modulation of the chemical potential
\begin{equation}
\mu(x) = \mu_0 (1 + A \cos(k x)).
\end{equation}
The $\vartheta$-term drives the instability towards formation of the staggered current order, accompanied by the modulation of a pseudoscalar condensate (dual to the axion field) and charge density
\begin{equation}
\mathrm{CDW:} \qquad J_y \sim J_y^{(1)}\sin(p x), \quad \Psi \sim \Psi^{(1)}\cos(p x), \quad \rho_{\mathrm{CDW}} \sim \rho_{\mathrm{CDW}}^{(0)} + \rho_{\mathrm{CDW}}^{(2)} \sin^2(p x),
\end{equation}
which are extracted as the subleading boundary coefficients of $A_y, \psi$ and $A_t$ bulk field profiles. The period of the charge modulation is twice smaller then the one of the staggered current. The presence of the homogeneous mode in $\rho(x)$ guarantees that the CDW produced in this way has a nonzero mean charge density, which appears as $\gamma$ coefficient in the hydrodynamic equations.
Note that the leading commensurate solutions are obtained with $p=1/2 \, k$ (1/1 commensurate CDW) and $p=1/1 \, k$ (2/1 commensurate CDW). In case when $p$ is close to the spontaneous momentum of the instability these solutions acquire additional stability due to lock in \cite{Andrade:2017leb,Andrade:2017ghg}.

\subsection{Thermodynamics}

In this appendix we discuss the holographic renormalization procedure \cite{deHaro:2000vlm} of our model, which allows us to extract the thermodynamic properties of our background solutions.
The UV expansions of the various fields are
\begin{align}
  Q_{tt} & = 1 + z^2 Q_{tt}^{(2)}(x) + z^3 Q_{tt}^{(3)}(x) + O(z^4) \\
  Q_{zz} & = 1 + z^2 Q_{zz}^{(2)}(x) + z^3 Q_{zz}^{(3)}(x) + O(z^4)  \\
   Q_{xx} & = 1 + z^2 Q_{xx}^{(2)}(x) + z^3 Q_{xx}^{(3)}(x) + O(z^4)  \\
   Q_{yy} & = 1 + z^2 Q_{yy}^{(2)}(x) + z^3 Q_{yy}^{(3)}(x) + O(z^4)  \\
  Q_{ty} & =  z^3 Q_{ty}^{(3)}(x) + O(z^4)  \\   
  Q_{zx} & =  z^3 Q_{zx}^{(3)}(x) + O(z^4)  \\   
  A_t &= \mu(x) - z \rho(x) + O(z^2) \\
  A_y & = z J_y(x) + O(z^2)  \\
  \psi &=  z^2 \psi^{(2)}(x) + O(z^3) 
\end{align}
%
%
From these expansions we can readily read-off the charge density $\rho(x)$ and the chemical potential $\mu(x)$. 
It is worth noting that the coefficients satisfy relations which correspond to the Ward identities $\langle T^\mu _\mu \rangle = 0$ and   $\partial_\mu \langle T^\mu _\nu \rangle = F^{\mu \nu} \langle J_\nu \rangle $. These are
\begin{align}
  Q_{tt}^{(3)}(x) + Q_{xx}^{(3)}(x) + Q_{yy}^{(3)}(x) &= 0 \\
 \partial_x Q_{xx}^{(3)} (x) + \frac{1}{2} \mu'(x) \rho(x) & = 0
\end{align}
The counter terms which renormalize the action are those written in \cite{Withers:2013loa}, 
\begin{equation}
  S_{ren} = S - \int d^3 x \sqrt{- h} (K - 4 + \psi^2)
\end{equation}
Therefore, we find that the components of the stress tensor read
\begin{align}
   \langle T_{tt} \rangle &= 2 + \frac{\mu_0^2}{2} - 3 Q_{tt}^{(3)}(x)  \equiv \epsilon(x) \\
    \langle T_{xx} \rangle&= 1 + \frac{\mu_0^2}{4} + 3 Q_{xx}^{(3)}(x)  \\
  \langle T_{yy} \rangle & = 1 + \frac{\mu_0^2}{4} + 3 Q_{yy}^{(3)}(x) \\
    \langle T_{ty} \rangle & = 3 Q_{ty}^{(3)}(x)
\end{align}
The entropy density is given by the area density of the black hole as
\begin{equation}
  s(x) = \frac{1}{4} \sqrt{Q_{xx}(x, 1) Q_{yy}(x, 1)}
\end{equation}
Equipped with these results, we can evaluate the thermodynamic quantities of our background solutions. As shown on the middle panel of Fig.\,\ref{fig:DCchecks}, the identity $\rho \mu + s T$ = $T_{xx} + T_{yy}$ (taking the spatial average on both sides) is satisfied in the solutions which we are considering.

%

%
%

\subsection{Conductivities} 

In order to evaluate the matrix of AC thermoelectric conductivities we consider the perturbative time dependent sources $\{\frac{\Delta_x T}{T}, E_x\}$ to the modes $\delta g_{tx}$ and $\delta A_x$, correspondingly. This leads, in DeDonder gauge \cite{Rangamani:2015hka} to the excitation of the total of 15 fields, coupled at the linear order. Technically, we find it quite important to introduce the sources explicitly by expanding the fluctuation profiles near the boundary. These expansions can be obtained by solving the equations of motion order by order at $z\rar 0$. In particular we use:
\begin{align}
\delta A_x(z,x,t) &= e^{-i \omega t} (1-z^4)^{-\frac{i \omega}{P(1)}} \left(E_x + \tilde{A}_x(z,x)\right), \qquad P(1)\equiv 3- \frac{\mu^2}{4} \\
\delta g_{tx}(z,x,t) &=  e^{-i \omega t}  \frac{1}{z^2}  (1-z^4)^{-\frac{i \omega}{P(1)}} \left(\frac{\Delta_x T}{T} + \frac{1}{6} \frac{\Delta_x T}{T} z^2 \omega^2 + z^2 \tilde{g}_{tx}(z,x)\right) \\
\delta g_{xz}(z,x,t) &=  e^{-i \omega t}  \frac{1}{f(z) z^2}  (1-z^4)^{-\frac{i \omega}{P(1)}} \left(\frac{i}{3} \frac{\Delta_x T}{T} z \omega + \frac{i}{6} \frac{\Delta_x T}{T} z^3 \omega^3 + z^3 \tilde{g}_{xz}(z,x)\right), 
\end{align}
and solve the linear perturbation equations for $\{\tilde{A}_x, \tilde{g}_{tx}, \tilde{g}_{xz}, \dots \}$. These redefinitions allow us to read off the expectation values as 
\begin{equation}
\la J_x \ra(x) = \p_z \tilde{A}_x(z,x) \Big|_{z=0}, \qquad \la T_{tx} \ra(x) =(-3) \p_z \tilde{g}_{tx}(z,x) \Big|_{z=0}.
\end{equation}
In order to get rid of the contribution of the contact terms we subtract the real part of the correlator at zero frequency from the AC results \cite{Kim:2014bza}. As a nontrivial check of our calculations, we get an excellent match between $\la J_x T_{tx} \ra (\omega)$ and $\la T_{tx} J_x \ra (\omega)$ correlators, which guarantees $\alpha = \bar \alpha$ in (6) in the main text

As mentioned in the main text, the DC conductivities can be evaluated without solving extra numerical equations of motion. In case of DC transport one can figure out the integrals of motion which allow to solve the radial evolution of the perturbation equations analytically and the problem reduces to solving a set of linear, elliptic equations (Stoke's equations) involving horizon data of the background solutions. In the particular case of 1D holographic lattices, this can be done by a series of spatial integrations so that we can arrive at closed, yet cumbersome, expressions. 
For our specific model, the general 1D expressions can be found in \cite{Donos:2017mhp}. 
This allows us to expand our analysis to lower temperatures, where the AC methods are too demanding.

\section{\label{app:fitting}Fitting of AC peaks}
The AC conductivities are obtained as a series of data points at various frequencies $\omega$. We fit this data in two stages. First we expand the data symmetrically to the negative frequencies and then find the maximum of the peak at $\omega=\hat{\omega}_0$. We fit the data with a pair of symmetric Lorentzians centered at $\hat{\omega}_0$ and $-\hat{\omega}_0$ as well as a constant background. More concretely, we fit the data to 
\begin{equation}
W_0(\omega) = C_0 + \frac{A_0}{(\omega-\hat{\omega}_0)^2 + \Sigma_0^2} + \frac{A_0}{(\omega+\hat{\omega}_0)^2 + \Sigma_0^2} 
\end{equation}
This gives a first approximation to the position, the width and the height of the peaks $\hat{\omega}_0$, $\Sigma_0$, and $A_0$.
As a second stage we fit the data with the ansatz function, which allows a leading order $\omega$-dependence in the numerator, taking into account the asymmetry of the peaks. We use the results of the previous fit as the seed parameters. We fit the parameters to the model
\begin{equation}
\label{eq:W1}
W_1(\omega) = C_1 + \frac{A_1 - B_1 (\omega^2 -\hat{\omega}_1^2)}{(\omega^2 -\hat{\omega}_1^2)^2 + \omega^2 \Sigma_1^2}
\end{equation}

Now take a look at the expressions for the AC conductivities of the main text. 
In the regime of weak pinning, we can consider $\Gamma$, $\Omega$ and $\omega_0$ as small parameters, as compared to the thermodynamic quantities and hydrodynamic coefficients, which are finite even without pinning. The exact scalings are model dependent, but for the case of $1/2$ lock-in phase, which we consider in the text, the scaling is $\Gamma\sim \lambda^2$, while $\Omega \sim \lambda$ and $\omega_0 \sim \lambda^{1/2}$ (where $\lambda\sim A$ is the scale of explicit symmetry breaking\cite{Andrade:2020hpu}). We can now expand all the coefficients in front of $\omega$, leaving only the leading terms: 
\begin{align}
\notag
\sigma(\omega)\! &= \! \sigma_0 \! + \! \frac{(\tilde{\rho}^2 \Omega  - 2 \tilde{\rho} \tilde{\gamma} \omega_0^2) - i \omega \tilde{\rho}^2}{\mu_0^2 \chi_{\pi \pi} ( - \omega^2 - i \omega \Omega + \omega_0^2) } \\
\notag
\frac{T}{\mu_0} \alpha(\omega)\! &= \! - \! \sigma_0 \! + \! \frac{(\tilde{\rho} \tilde{s} \Omega  - (\tilde{s} - \tilde{\rho}) \tilde{\gamma} \omega_0^2) - i \omega \tilde{\rho} \tilde{s}}{\mu_0^2 \chi_{\pi \pi} ( - \omega^2 - i \omega \Omega + \omega_0^2) } \\
\label{equ:AC_formulae_small_lambda}
\frac{T}{\mu_0^2} \bar{\kappa}(\omega)\! &= \! \sigma_0 \! + \! \frac{(\tilde{s}^2 \Omega  + 2 \tilde{s} \tilde{\gamma} \omega_0^2 )- i \omega \tilde{s}^2}{\mu_0^2 \chi_{\pi \pi} ( - \omega^2 - i \omega \Omega + \omega_0^2)  }, 
\end{align}
The linear frequency dependence of the numerator in these expressions leads to an asymmetric peak of the form \eqref{eq:W1} in the real part of the conductivity. More precisely (and taking into account the definitions of tilde-parameters from the main text), the corresponding values of the fitting parameters are 
\begin{align}
\notag
\sigma:& &  C_1&= \sigma_0, &  A_1 &= \frac{\rho^2}{\chi_{\pi \pi}} \Omega \omega_0^2, &   B_1 &= -2 \frac{\rho}{\chi_{\pi \pi}} \Omega \gamma, &   \Sigma_1 &=\Omega, &   \hat{\omega}_1 = \omega_0 \\
\notag
\frac{T}{\mu} \alpha:& &  C_1&= -\sigma_0, &   A_1 &= \frac{s T \rho}{\mu \chi_{\pi \pi}} \Omega \omega_0^2, &   B_1 &= \frac{\mu \rho - s T}{\mu \chi_{\pi \pi} } \Omega \gamma, &   \Sigma_1 &=\Omega, &   \hat{\omega}_1 = \omega_0 \\
\label{equ:fit_params}
\frac{T}{\mu^2} \kappa:& &  C_1&= \sigma_0, &   A_1 &= \frac{(s T)^2}{\mu^2 \chi_{\pi \pi}}\Omega \omega_0^2, &   B_1 &=  2 \frac{s T}{\mu \chi_{\pi \pi}} \Omega \gamma, &  \Sigma_1 &=\Omega, &   \hat{\omega}_1 = \omega_0 
\end{align}
We see that from each fit we can obtain $\sigma_0, \Omega$ and $\omega_0$ directly and independently. As we show on Fig.\,\ref{fig:ACchecks} (left panel), the parameters obtained this way match perfectly with each other. As another, more nontrivial consistency check, we can extract certain thermodynamic quantities (see \eqref{equ:fit_params}) from the fit of $A_1$, using $\Omega$ and $\omega_0$ and compare them with the same quantities extracted as the expectation values of the corresponding operators for the background solutions. We plot the ratio of these differently obtained thermodynamic quantities ($\hat{A}_1$) on the right panel of Fig.\,\ref{fig:ACchecks} and see that it matches unity within a few percent margin. Finally, the $\gamma$-coefficient is related to the $B_1$ -- the peak asymmetry parameter, which has the highest numerical uncertainty.\footnote{In the vicinity of the phase transition the analysis of the quasinormal modes, performed in \cite{Andrade:2020hpu} gives a better control of the physics. We are not considering this regime here.} Nonetheless, it shows a very good agreement in in the regime of larger $c_1$, where the peaks are wider and the quality of fitting improves, this is shown on the right panel of Fig.\,\ref{fig:ACchecks} as well.  

\begin{figure}
\includegraphics[width=0.45 \linewidth]{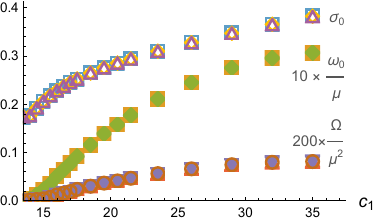} \ \
\includegraphics[width=0.45 \linewidth]{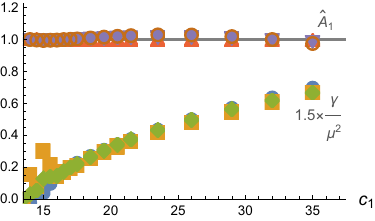}
\caption{\label{fig:ACchecks} The fitting parameters, as obtained from the independent fits of AC electric and thermal conductivities as well as thermopower with the function \eqref{eq:W1}. \textbf{The left panel} shows that the bare parameters match perfectly well between the fits.\textbf{ The right panel} shows the agreement between the thermodynamic quantities obtained from the AC fits and evaluated on the background solutions. It also demonstrates the lesser accuracy in $\gamma$-parameter in the regime of weaker order parameter.}
\end{figure}

Another important cross check of our results is the comparison with the DC conductivities, which can be evaluated given only the near horizon data of the background solutions. We plug in the parameters obtained from the AC fits \eqref{equ:fit_params} to the expressions for DC conductivities listed in the main text and compare this to the DC conductivities evaluated directly from the horizon data. As we show on Fig.\,\ref{fig:DCchecks}, they agree perfectly well.

\begin{figure}
\includegraphics[width=0.3 \linewidth]{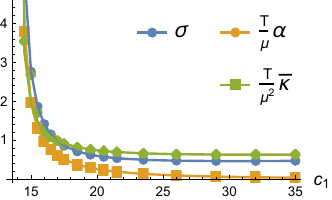}\ \
\includegraphics[width=0.3 \linewidth]{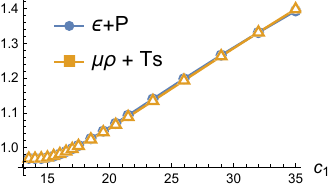} \
\includegraphics[width=0.35 \linewidth]{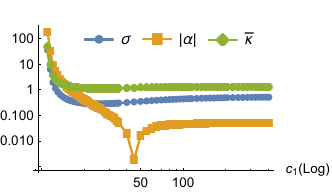}
\caption{\label{fig:DCchecks} \textbf{Left panel:} Comparison between the DC conductivities obtained directly from the horizon data and those evaluated with eq.(9) in the main text using the hydrodynamic parameters extracted from AC fits. \textbf{Middle panel:} Check of the thermodynamic identity in the pseries of periodic backgrounds. \textbf{Right panel:} Evolution of DC conductivities in periodic model at large $c_1$. (Log-Log scale). This continues the data shown on the left panel. The change of sign in $\alpha$ is evident here at strong order regime, similarly to the behavior at small temperature shown on Fig.\,2 in the main text.}
\end{figure}

\begin{figure*}[t]
\includegraphics[width=0.32 \linewidth]{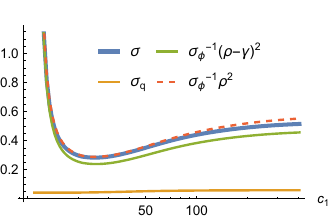}
\includegraphics[width=0.32 \linewidth]{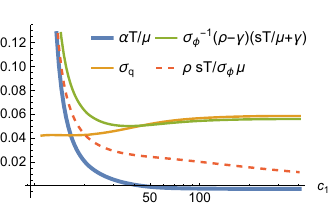}
\includegraphics[width=0.32 \linewidth]{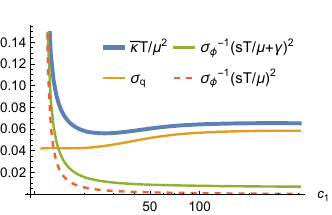}
\caption{\label{fig:Cseries_contrib} The contributions to DC conductivities from various terms in eq.(9) in the main text in a series of coupling $c_1$ in the periodic model. At high $c_1$, the electric conductivity is dominated by the $\sigma_\phi^{-1}$ term, the heat conductivity -- by $\sigma_q$-term, while the thermopower is a result of a fine balance between them. Red lines assume $\gamma=0$ and help to appreciate the role of $\gamma$. Data as on Fig.\,\ref{fig:DCchecks} for $T/\mu=0.1$.}
\end{figure*}

%

\section{\label{sec:helix} Helical Bianchy VII holographic model}

A vast literature is available on the holographic models which realize translation symmetry breaking in a homogeneous way \cite{Andrade:2013gsa,Donos:2013eha,Baggioli:2014roa,Alberte:2015isw,Amoretti:2018tzw,Ammon:2019wci,Donos:2019hpp,Amoretti:2019cef,Baggioli:2022pyb,Baggioli:2021xuv}, but they lack dynamical mechanism for pattern formation. We won't discuss this approach here since we will among other things, be interested in the physics of growing order parameter. Another example of dynamical pattern formation in holography has been studied in connection to helical orders \cite{Nakamura:2009tf,Ooguri:2010kt,Donos:2012wi,Donos:2012js,Donos:2014oha,Andrade:2018gqk} and we  discuss it here.

\noindent We have checked that $\alpha$ also changes sign as we vary the temperature in the Bianchy VII solutions that break translations homogeneously. Our setup and notation closely follow \cite{Andrade:2018gqk, Andrade:2020hpu}.
The model is defined in 5-dimensional bulk ($x^\mu = \{t,x,y,z,u\}$, $u$-- radial holographic coordinate, with boundary at $u=0$) with dynamical gravity, an Abelian gauge field $A_\mu$, dual to the chemical potential and an auxiliary vector field $B_\mu$, which we use to source the explicit translational symmetry breaking. The action reads
\begin{equation}
\label{eq:action_helix}
  S = \int d^5 x \sqrt{- g} \left( R  - 2 \Lambda - \frac{1}{4} F^2 - \frac{1}{4} W^2  \right) - 
   \frac{\bm{\gamma_{CS} }}{6} \int  A \wedge F \wedge F ,
\end{equation} 
where $\Lambda=-6$ and $F \equiv dA$, $W\equiv dB$ -- the field strength tensors.  As shown in \cite{Ooguri:2010kt, Donos:2012wi} the model in absence of explicit breaking source develops an instability at a $\gamma_{CS}$-dependent critical temperature, which leads to a formation of the structure characterized by the helical forms
\begin{align}
\label{equ:helical_forms}
\omega^{(k)}_1 & = dx \\
\notag
\omega^{(k)}_2 & = \cos (k x) dy - \sin(k x) dz \\
\notag
\omega^{(k)}_3 & = \sin (k x) dy + \cos(k x) dz. 
\end{align}
We break translations explicitly preserving the shape of the helix, by introducing the field $B_\mu$, with boundary condition
\begin{equation}
\label{eq:bc B2}
B\big|_{u\rar0} = \lambda \omega^{(k)}_2,
\end{equation}
\noindent 
so that the conductivity matrix is finite at zero frequency \cite{Donos:2012js, Donos:2014oha}. 
We take the ansatz 
\begin{align}
\label{ansatz helix 1}
  ds^2 &= u^{-2} [ - T f  dt^2 + U/f du^2 + W_1 (\omega^{(k)}_1)^2 + W_2 (\omega^{(k)}_2 + Q dt )^2 + W_3 (\omega^{(k)}_3)^2] \\
  A &= A_t dt + A_2 \omega^{(k)}_2 \\
 \label{ansatz helix 3}
  B &= B_t dt + B_2 \omega^{(k)}_2
\end{align} 
\noindent where 
\begin{equation}
  f =  (1-u^2) (1 + u^2 -\mu ^2 u^4/3 )
\end{equation}
All unknowns are functions of the radial coordinate $u$ only. We impose the DeTurk gauge as in \cite{Andrade:2017cnc}. 
The normal phase is the Schwarzschild AdS solution with 
\begin{align}
  T &= U = W_i = 1, & Q &= 0,  & A_t &= \mu (1- u^2), & A_2 &= B_t = B_2 = 0.
\end{align} 
At $\bm{\gamma_{CS}} = 3$, the marginal mode of highest temperature occurs at \cite{Andrade:2020hpu}
\begin{equation}
\label{equ:helix_preferred}
  k/\mu = 2.18, \qquad (T/\mu)_c = 0.223
\end{equation}
Taking this as a starting point, we turn on $\lambda > 0$ and build a various families of solutions which break translations spontaneously and explicitly for increasing $\bm{\gamma_{CS}}$. 
The DC conductivities in terms of horizon data are given in Appendix C of \cite{Andrade:2018gqk}. Performing the same analysis as for the periodic model in the main text, we arrive at the results summarized on Figs.\,\ref{fig:DCcond_helix},\ref{fig:EFT_vs_T_helix}, \ref{fig:EFT_contrib_helix}, which are in a perfect qualitative agreement with the periodic case.

\begin{figure*}[ht]
\includegraphics[width=0.32 \linewidth]{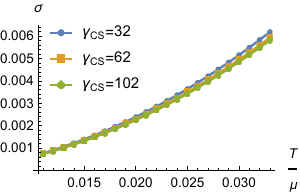} \
\includegraphics[width=0.32 \linewidth]{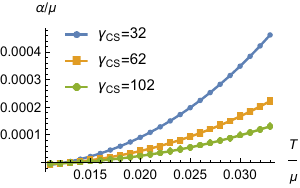} \ 
\includegraphics[width=0.32 \linewidth]{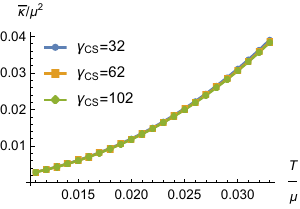} \ 
\caption{\label{fig:DCcond_helix} The evolution DC thermo-electric conductivities at small temperatures. The holographic helical model with different values of the coupling $\gamma_{CS}$ is considered. Data taken for $k/\mu = 0.76$, $\lambda/\mu = 0.3$.}
\end{figure*}

\begin{figure*}[ht]
\includegraphics[width=0.32 \linewidth]{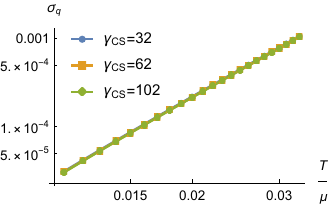} \
\includegraphics[width=0.32 \linewidth]{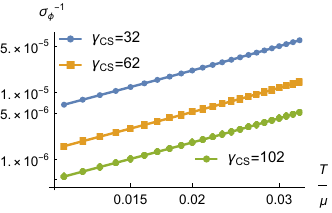} \
\includegraphics[width=0.32 \linewidth]{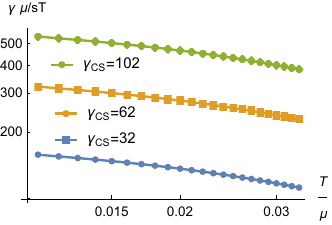}
\caption{\label{fig:EFT_vs_T_helix} Behavior of the effective theory parameters at low temperatures in the holographic helical model in Log-Log scales. The incoherent conductivity $\sigma_q$ and Goldstone diffusivity behave as certain power laws, but the former does not depend on the coupling $\gamma_{CS}$. Same data as on Fig.\,\ref{fig:DCcond_helix}.}
\end{figure*}

\begin{figure*}[ht]
\includegraphics[width=0.32 \linewidth]{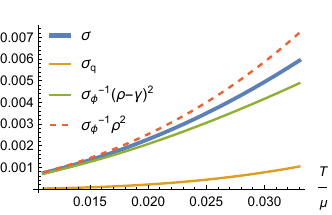} \
\includegraphics[width=0.32 \linewidth]{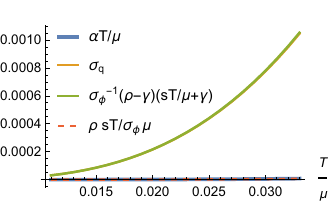} \
\includegraphics[width=0.32 \linewidth]{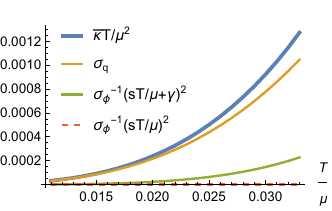}
\caption{\label{fig:EFT_contrib_helix} Contributions to DC conductivities from various terms in eq.\,(9) in the main text in the holographic helical model. At low $T$, the electric conductivity is dominated by $\sigma_\phi^{-1}$ term, the heat conductivity -- by the $\sigma_q$-term, while thermopower is a result of a fine balance between them. This is the same behavior as for the periodic model in the main text. Red lines assume $\gamma=0$ and help to appreciate the role of $\gamma$. Data as on Fig.\,\ref{fig:DCcond_helix} for $\gamma_{CS}=62$.}
\end{figure*}

\section{\label{sec:ebound} Checks of the entropy bound}

In \cite{Amoretti:2019cef,Amoretti:2018tzw} it has been suggested that at low temperatures the Goldstone diffusivity is saturated by the entropy current, which leads to the universal relaxation bound (in our notation, see the correspondence in \cite{Armas:2020bmo}) 
\begin{equation}
\label{equ:ebound}
\frac{\gamma}{\sigma_\phi} \sim \frac{\mu}{sT}  \left(\sigma_q + \frac{\gamma^2}{\sigma_\phi}\right)
\end{equation}
On the other hand, the positivity of the entropy production requires $\sigma_q \geq 0$ and in case it saturates the relation \eqref{equ:ebound} simplifies to $\gamma \mu / s T \sim 1$. Note that this is the expression we plot on Fig.\,4 of the main text and Fig.\,5 here and that in our examples it is not quite obeyed. 

The saturation of both the relaxation bound \eqref{equ:ebound} as well as the entropy production bound $\sigma_q \geq 0$ has been observed in the homogeneous Q-lattice holographic model in \cite{Amoretti:2019cef,Amoretti:2018tzw}. Here we demonstrate how the bound \eqref{equ:ebound} comes about in our data. As one can see on Fig.\,\ref{fig:ebound}, there are no signs of saturation in our examples. The possible reasons for that might be the presence of $\theta$-term and Chern-Simons term in our models, which has not been considered in \cite{Amoretti:2019cef,Amoretti:2018tzw}, or the fact that we don't reach low enough temperature. Otherwise, the absence of saturation here may simply point out that the above mechanism is model dependent and may or may not dominate the Goldstone relaxation in different setups. 

\begin{figure*}[ht]
\includegraphics[width=0.32 \linewidth]{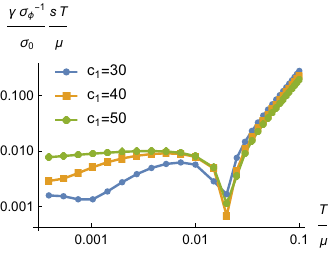} 
\includegraphics[width=0.32 \linewidth]{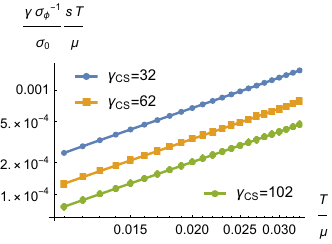} 
\includegraphics[width=0.32 \linewidth]{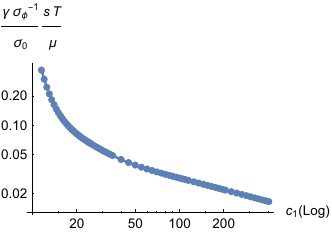}
\caption{\label{fig:ebound} Checks of the universal relaxation bound \eqref{equ:ebound} suggested in \cite{Amoretti:2019cef,Amoretti:2018tzw}, note $\sigma_0 \equiv \sigma_q + \gamma^2/\sigma_\phi$.\textbf{ Left panel:} $T$ series in the periodic model, data as on Fig.\,2 in the main text. \textbf{Middle panel:} $T$ series in helical model, data as on Fig.\,\ref{fig:DCcond_helix}. \textbf{Right panel:} $c_1$ series in the periodic model, data as on Fig.\,\ref{fig:Cseries_contrib}. We do not observe the saturation of the bound ($\gamma s T / \sigma_0 \sigma_\phi \mu \approx 1$) in any of the cases.}
\end{figure*}



\end{widetext}

\bibliography{strange_Mott}

\end{document}